\documentclass[a4paper]{article}
\usepackage{microtype}
\usepackage{url}
\usepackage{color}
\usepackage[colorlinks,linkcolor=black,urlcolor=black,citecolor=black]{hyperref}
\usepackage{INTERSPEECH2022}
\usepackage{amsmath}
\usepackage{siunitx}
\usepackage{amsmath,amsfonts,bm}

\def\eqref#1{equation~\ref{#1}}
\def\1{\bm{1}}

\DeclareMathAlphabet{\mathsfit}{\encodingdefault}{\sfdefault}{m}{sl}
\SetMathAlphabet{\mathsfit}{bold}{\encodingdefault}{\sfdefault}{bx}{n}

\usepackage{soul}

\title{Separate What You Describe: Language-Queried Audio Source Separation}
\name{Xubo Liu$^1$, Haohe Liu$^1$, Qiuqiang Kong$^2$, Xinhao Mei$^1$, Jinzheng Zhao$^1$, \\Qiushi Huang$^1$, Mark D. Plumbley$^1$, Wenwu Wang$^1$}
\address{
  $^1$School of Computer Science and Electronic Engineering, University of Surrey, UK\\
  $^2$Speech, Audio, and Music Intelligence (SAMI) Group, ByteDance, China
  }
\email{\{xubo.liu, hl01486, m.plumbley, w.wang\}@surrey.ac.uk, kongqiuqiang@bytedance.com}

\begin{document}

\maketitle
\begin{abstract}
    % Previous work on source separation mainly focused on separating sound mixtures into specific sources belonging to a restricted domain of source classes, such as vocal and music. 
    % In existing informed source separate systems, target audio sources are separated from a mixture, with class label or embedding of the target source given as the query information. In this paper, we introduce a novel method, i.e. language-queried audio source separation (LASS), which aims to separate the target source from an audio mixture using natural language description as a query (e.g., ``\textit{a man tells a joke followed by people laughing}''). We propose LASS-Net which consists of two parts: (1) a query network to extract the embedding from the language description based query; (2) a separation network that takes mixture and query embedding as input and extracts the target source. 
    In this paper, we introduce the task of language-queried audio source separation (LASS), which aims to separate a target source from an audio mixture based on a natural language query of the target source (e.g., ``\textit{a man tells a joke followed by people laughing}''). A unique challenge in LASS is associated with the complexity of natural language description and its relation with the audio sources. To address this issue, we proposed LASS-Net, an end-to-end neural network that is learned to jointly process acoustic and linguistic information, and separate the target source that is consistent with the language query from an audio mixture. We evaluate the performance of our proposed system with a dataset created from the AudioCaps dataset. Experimental results show that LASS-Net achieves considerable improvements over baseline methods. Furthermore, we observe that LASS-Net achieves promising generalization results when using diverse human-annotated descriptions as queries, indicating its potential use in real-world scenarios. The separated audio samples and source code are available at \url{https://liuxubo717.github.io/LASS-demopage}.
    
% To evaluate the performance of our model, we create a dataset GUESS (lan\textbf{gu}age-based targ\textbf{e}t \textbf{s}ound \textbf{s}eparation) based on the AudioCaps dataset. Our model achieves an average SDR of 6.4 dB on GUESS. 

% language-queried target sound separation is a challenging task, as it requires capturing sound event phrases (e.g., a person speaking) in an input query and understanding their relationships, and then separating sounds from one or more sources that are consistent with the language description. 

% We create a dataset GUESS-50K for lan\textbf{gu}age-based targ\textbf{e}t \textbf{s}ound \textbf{s}eparation. GUESS-50K consists of about 5000 audio clips drawn from the AudioCaps dataset for a total of 15 hours. Our model achieves an average SDR of 6.4 dB on GUESS-50K. 
\end{abstract}

\section{Introduction}
% For example, in a live concert, people can focus on the singer's voice and music, while ignoring the applause of the audience.
Human beings can focus their auditory attention on specific sounds in environments \cite{yang2021detect}. Source separation systems aim to separate mixtures of sound sources, which is the basis for computational auditory scene analysis \cite{wang2006computational}. Recently, significant progress has been made in audio source separation such as speech separation \cite{wang2018supervised, luo2019conv}, music source separation \cite{kong2021decoupling, liu2020channel, liu2021cws} and universal sound separation \cite{kong2020source, kavalerov2019universal, tzinis2020improving, wisdom2021s}.

% Meanwhile, in real-world application scenarios, users are usually interested in some specific sounds. 
% Instead of using a predefined fixed set of sound categories, natural language expression could be a more effective alternative.
One major challenge for existing source separation systems is to deal with a vast number of sound classes in the real world. When multiple sources are presented simultaneously, it is difficult for these systems to obtain accurate separation results \cite{tzinis2020improving, ochiai2020listen}. Previous methods \cite{kong2020source, ochiai2020listen} investigated using source category information as a query to separate a specific source from a mixture, as a way to reduce the difficulty of source separation. However, the information provided by source categories is often limited. In practical applications, instead of using a predefined fixed set of source categories, one may prefer to use a natural language description to identify and separate the target sound source. Such natural language descriptions can include auxiliary information for source separation such as spatial and temporal relationships of sound events, such as ``\textit{dog barks in the background}'' or ``\textit{people applaud followed by a woman speaking}''. To our knowledge, audio source separation with natural language queries has not been investigated in the literature.
% Existing source separation methods mainly focus on separating sounds belonging to a predefined fixed set of source classes, such as vocal and music. When people want to extract a sound clip containing multiple sources from a mixture, one possible solution is to apply a source separation method to extract each individual source component and remix them. However, this approach is time-consuming, especially for users without expert knowledge.

This paper introduces a language-queried audio source separation (LASS) task. Given an audio mixture and a natural language query of the target source, LASS aims to automatically extract the target source from the mixture, where the target source is consistent with the language query. LASS provides a potentially useful tool for future source separation systems, allowing users to extract desired audio sources via natural language instructions. Such a system could be useful in many applications, such as automatic audio editing \cite{rubin2013content}, multimedia content retrieval \cite{oncescu2021audio}, and controllable hearable devices \cite{ochiai2020listen}. 

The challenges of achieving LASS are associated with the complexity of natural language expressions and the characterization of their relation with sound sources. The language description of a sound source usually consists of multiple phrases (e.g., ``\textit{people speak and music plays}''), each phrase referring to a sound event in the audio mixture. In addition, the same audio source can be delivered with diverse language expressions, such as ``\textit{music is being played with a rhythmic beat}'' or ``\textit{an upbeat music melody is playing over and over again}''. In summary, LASS not only requires these phrases and their relationships to be captured in the language description, but also one or more sound sources that match the language query should be separated from the audio mixture.

% A unique challenge in LASS is the complexity of natural language expressions and their correlation with sound sources. A language description usually consists of multiple sound event phrases (e.g., ``\textit{a person speaking}'', ``\textit{typing}''), each referring one sound event in the audio mixture. In addition, an audio source can be presented with diverse expressions, such as ``\textit{something goes round that is playing its song}'' and ``\textit{fair kind of music is being played}''. LASS not only requires to capture these phrases and their relationships in the language description, but also to separate one or more sounds that match the language query from the mixture. 
% Different from them, LASS using open natural language as queries which can provide side information useful for source separation, such as spatial and temporal relationships of acoustic events. LASS is a challenging task as it not only requires capturing sound event phrases (e.g., ``\textit{a person speaking}'', ``\textit{typing}'') in language descriptions, but also separating one or multiple sources that match with the language queries. In addition, an audio source can be presented with diverse language descriptions, such as ``\textit{something goes round that is playing its song}'' and ``\textit{fair kind of music is being played}'', which poses a further challenge to LASS. 

In this work, we present LASS-Net, which is trained to jointly process acoustic and linguistic information, and separate the target source described by the natural language expressions. In LASS-Net, a Transformer-based \cite{devlin2018bert} query network is used to encode the language expression into a query embedding, and a ResUNet-based \cite{kong2021decoupling} separation network is then used to separate the target source from mixture conditioned on the query embedding. To evaluate the performance of our model, we create a dataset based on the AudioCaps \cite{kim2019audiocaps} dataset, experimental results demonstrate that our model can achieve considerable improvements over baseline methods. We also observe that LASS-Net shows promising separation performance when queried by diverse human-annotated descriptions, indicating the potential for generalization in real application scenarios.

The remainder of this paper is organized as follows. We review the related work in Section \ref{sec-2}. The LASS-Net is presented in Section \ref{sec-3}. We describe the dataset we created in Section \ref{sec-4}, and experiments and results in \ref{sec-5}. We draw our conclusions in Section \ref{sec-6}, together with discussions on future work.

\begin{figure*}
  \centering
  \includegraphics[width=\linewidth]{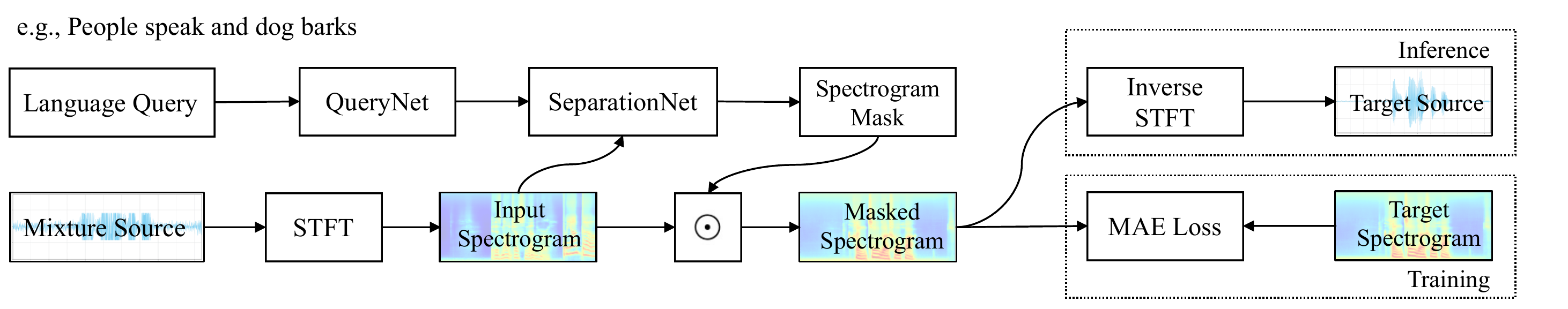}
  \caption{Framework of our proposed LASS-Net.}
  \label{fig-1}  
%   \vspace{-1.25em}
\end{figure*}

\section{Related Work}
\label{sec-2}
Our work relates to several tasks in the literature: universal sound separation, target source extraction, and audio captioning. We will discuss each of these as follows.

\subsection{Universal sound separation}
Universal sound separation (USS) \cite{kavalerov2019universal} is the task of separating a mixture of arbitrary sound sources in terms of their classes. USS is a challenging problem since the number of sound classes in the real world is very large. Several approaches have been proposed to address the issue of a large number of sound classes, such as leveraging semantic information learned by sound classifiers \cite{tzinis2020improving} and establishing large-scale high-quality datasets \cite{wisdom2021s}. In a similar way to USS, LASS aims to separate real-world sounds, but the objective of LASS is to perform the separation using natural language descriptions as queries.

\subsection{Target source extraction}
Target source extraction (TSE) aims to separate a specific source from an audio mixture given query information such as a sound event tag. In contrast to USS, TSE only extracts sources of interest from the mixture. There are several applications for this problem, such as target speech extraction using speaker information \cite{wang2018voicefilter, wang2020voicefilter}, and target sound extraction using acoustic event tags \cite{ochiai2020listen} or onomatopoeic words \cite{okamoto2021environmental}. In contrast to TSE, LASS focuses on extracting target audio sources that match linguistic queries.

% When multiple sounds are presented in an audio mixture simultaneously, it is difficult for existing source separation systems to obtain accurate separation results. In real-world application scenarios, users usually only interest in some specific sounds. Similarly, target sound extraction aims to separate some specific sounds from mixtures, which potentially reduces the difficulty of source separation. 

% environmental sound extraction using onomatopoeic words \cite{okamoto2021environmental}, and singing voice extraction using humming \cite{smaragdis2009separation}

\subsection{Automated audio captioning}
Automated Audio captioning (AAC) \cite{drossos2017automated, mei2021encoder, liu2021cl4ac, mei2021diverse} is the task where natural language descriptions are generated for an audio clip. Recently, AAC has attracted increasing interest in the Detection and Classification of Acoustic Scenes and Events (DCASE) community. There are several datasets for AAC, such as Clotho \cite{drossos2020clotho} and AudioCaps \cite{kim2019audiocaps}. AudioCaps is the largest public dataset for AAC research and has been used in several recent studies \cite{mei2021audio, liu2022leveraging}. In this work, we create a dataset for LASS based on the AudioCaps dataset. This new dataset will be described in Section \ref{sec-4}.

\section{Proposed Approach} 
\label{sec-3}
We propose LASS-Net, a neural network which separates target audio sources with natural language queries, as shown in Figure \ref{fig-1}. LASS-Net consists of two components: a query network (QueryNet) that takes a language query as input and outputs a query embedding, and a separation network (SeparationNet) that takes mixture and query embedding as input and predicts the target source. These two modules are trained jointly. We describe the details of LASS-Net in the following sections.

% a separation network that takes mixture and query embedding as input and predicts the target source. 

\subsection{LASS-Net}
% Regression-based approach has been used in recent studies for source separation \cite{kong2020source, kavalerov2019universal, tzinis2020improving, wisdom2021s}, where a regression function $f(\cdot)$ is learned to map an audio mixture $x$ to a target source $s$:
% \begin{equation}
%     \label{eq-1}
%     f(x) \mapsto s.
% \end{equation}
In LASS-Net, We build the separation in the time-frequency domain. We denote the language query as $q$, and deploy a query network to extract a query embedding $e_q$:
\begin{equation}
    \label{eq-3}
    \operatorname{QueryNet}(q) \mapsto e_q.
\end{equation}
The audio mixture $x$ is transformed to the spectrogram $X$ using short-time Fourier transform (STFT). The magnitude spectrogram and phase of $X$ are denoted as $|X|$ and $e^{j{\angle}X}$, where $X = |X|e^{j{\angle}X}$. Magnitude spectrogram $|X| \in \mathbb{R}^{F \times T}$ is a two-dimensional time-frequency representation, where $T$ and $F$ represent the number of time frames and the dimension of the spectral feature, respectively. Our objective is to learn a regression from $|X|$ to the magnitude spectrogram $|\hat{S}|$ of the estimated target source $\hat{s}$, conditioned on the query embedding $e_q$. Specifically, we use a separation network that accepts inputs $|X|$ and $e_q$, and outputs a latent feature $Z$ with the same shape as $|X|$:
\begin{equation}
    \label{eq-2}
    \operatorname{SeparationNet}(|X|, e_q) \mapsto Z \in \mathbb{R}^{F \times T}.
\end{equation}
Then, $Z$ is fed into an element-wise sigmoid function $\sigma(\cdot)$ to obtain a spectrogram mask $M$:
\begin{equation}
    \label{eq-2}
    \sigma(Z) = M \in [0, 1]^{F \times T}.
\end{equation}
The magnitude spectrogram $|\hat{S}|$ of the estimated target source is obtained by masking $|X|$:
\begin{equation}
    \label{eq-5}
    |\hat{S}| = M \odot |X| \in \mathbb{R}^{F \times T},
\end{equation}
where $\odot$ is the Hadamard product. 

The training objective is to minimize the mean absolute error (MAE) between $|\hat{S}|$ and magnitude spectrogram $|S|\in \mathbb{R}^{F \times T}$ of the ground truth target audio source. The MAE loss term is defined as follows:
\begin{equation}
    \label{eq-6}
    \operatorname{Loss_{MAE}} = \left\lVert |S|-\hat{|S|} \right\lVert_1,
\end{equation}
where $\left\lVert.\right\lVert$ is an $l_1$ norm.

The mixture phase $e^{j{\angle}X}$ is reused to recover the STFT spectrogram $\hat{S}$ from the estimated magnitude spectrogram $|\hat{S}|$, where $\hat{S} = |\hat{S}|e^{j{\angle}X}$. Finally, inverse STFT is applied on $\hat{S}$ to obtain the estimated source $\hat{s}$.

\subsection{Query network}
To extract the language query embedding, we use BERT \cite{devlin2018bert} as a query network in the proposed LASS-Net. BERT is a language model pre-trained on large-scale text datasets (e.g., BooksCorpus \cite{zhu2015aligning}). BERT contains prior linguistic knowledge such as syntactic and semantic information, which is useful in audio-language tasks, as shown in audio captioning \cite{liu2022leveraging, weck2021evaluating}.  Concretely, we use the pre-trained BERT \cite{turc2019well} consisting of \num{4} Transformer encoder blocks, each with \num{4} head and \num{256} hidden dimensions, respectively. The input language query $q=\{q_n\}_{n=1}^N$ which consists of $N$ words is fed into the BERT model, as a result, a 256-dimensional word-level embedding $e=\{e_n\}_{n=1}^N$ is obtained. In the BERT model, the bidirectional self-attention is used to consider both past and future context, therefore, we adopt the first embedding $e_1$ as the output embedding of BERT. Finally, $e_1$ is passed to a fully-connected layer with \num{256} nodes and ReLU activation to obtain the language query embedding $e_q$.
\label{section-3.2}
\subsection{Separation network}
We design the separation network based on ResUNet \cite{kong2021decoupling}, which is an improved UNet model used previously in \cite{kong2021decoupling, liu2020channel, kong2020source} for source separation. The ResUNet consists of six encoder blocks and six decoder blocks. There are skip connections between encoder and decoder blocks at the same level. The encoder and decoder blocks share the same structure that contains two ConvBlocks. Each ConvBlock consists of a batch normalization, a leakyReLU activation, and a convolutional layer with kernel size $4 \times 4$. In the encoder blocks, average pooling is applied for downsampling. In the decoder blocks, transpose convolution is applied for upsampling. The number of feature maps of each encoder block are \num{32}, \num{64}, \num{128}, \num{256}, \num{384}, and \num{384}, respectively, and the number of feature maps of each decoder block is \num{384}, \num{384}, \num{256}, \num{128}, \num{64}, and \num{32}, respectively. After the last decoder block, a \num{32}-channel ConvBlock followed by a $1 \times 1$ convolutional layer is deployed to estimate the spectrogram mask, which has the same shape as the input spectrogram.

To bridge the query network and the separation network, we use the Feature-wise Linearly modulated (FiLm) layer \cite{perez2018film} after each ConvBlock deployed in the separation network. Specifically, let $H^{(l)} \in \mathbb{R}^{m \times h \times w}$ denote the output feature map of a ConvBlock $l$ that has $m$ filters. The modulation parameters are applied per feature map $H^{(l)}_i$ with the FiLm layer as follows:
\begin{equation}
    \label{eq-3}
    \operatorname{FiLM}(H^{(l)}_i|\gamma^{(l)}_i, \beta^{(l)}_i) = \gamma^{(l)}_iH^{(l)}_i + \beta^{(l)}_i,
\end{equation}
where $H^{(l)}_i \in \mathbb{R}^{h \times w}$, and $\gamma^{(l)}, \beta^{(l)} \in \mathbb{R}^{m}$ are the modulation parameters from $g(.)$, i.e., $(\gamma, \beta)=g(e_q)$, such that $g(.)$ is a neural network and $e_q$ is the language embedding obtained from the query network. In this work, we model $g(\cdot)$ with two fully connected layers followed by ReLU activation, which is jointly trained with the separation network.
\section{Dataset}
\label{sec-4}
We have created a dataset for LASS based on the AudioCaps \cite{kim2019audiocaps} dataset. AudioCaps is the largest publicly available dataset for audio-language research, containing approximately \num{50}k \num{10}-second audio clips from AudioSet \cite{gemmeke2017audio}. AudioCaps provides one human-annotated caption for each audio clip in the training set, and we can get sound event tags for each audio clip from AudioSet. Note that there is no direct connection between these two annotations. To ensure the diversity of audio clips, we first select \num{33} sound event tags (e.g., speech, typing, vibration, dog, rain) that belong to five root categories of AudioSet Ontology\footnote{\url{https://research.google.com/audioset/ontology/index.html}}: Human sounds, Animal, Sound of things, Natural sound, and Channel, environment and background. Then, we retrieve audio clips from AudioCaps if their sound event tags are all contained in these \num{33} sound event classes. As a result, \num{6244} audio clips ($\sim$\num{17.3} hours) are retrieved, each audio clip has one caption in AudioCaps. We divide these audio clips into training data and test data of \num{6044} and \num{200} audio clips, respectively.

To create the audio mixtures, we first select an audio clip as the target source, and then randomly select an audio clip as the background source, whose sound event tag does not overlap with that of the target source. We mix the target source and the background source with a signal-to-distortion ratio (SDR) at 0 dB. The training mixtures are created on-the-fly using training data. To create the test mixtures, each test audio is mixed with five randomly selected background sources. As a result, \num{1000} test mixtures are created, each mixture has a ground truth target source audio and one language query, we denote these test mixtures as LASS-Test.

In practice, human descriptions of an audio source are often diverse. To evaluate the model performance under diverse language expressions, we first randomly select \num{50} audio clips from \num{200} test audio and invite five language experts to annotate them. Each person labels one description per audio clip without any hinters and restrictions. As a result, we collect an audio test subset, where each audio has six captions (one from AudioCaps and five from the annotators we recruited). Then, the corresponding \num{250} test audio mixtures are drawn from the LASS-Test to create a test subset mixtures, referred to LASS-Sub-Test.
\section{Experiments and Results}
\label{sec-5}

% \begin{equation}
%     \label{eq-3}
%     \operatorname{SNR}(s_{target}, s_{background}) = 10 \log_{10}(\frac{\left\lVert s_{target} \right\lVert^2} {\left\lVert s_{background} \right\lVert^2})
% \end{equation}

\begin{figure*}
  \centering
  \includegraphics[width=\linewidth]{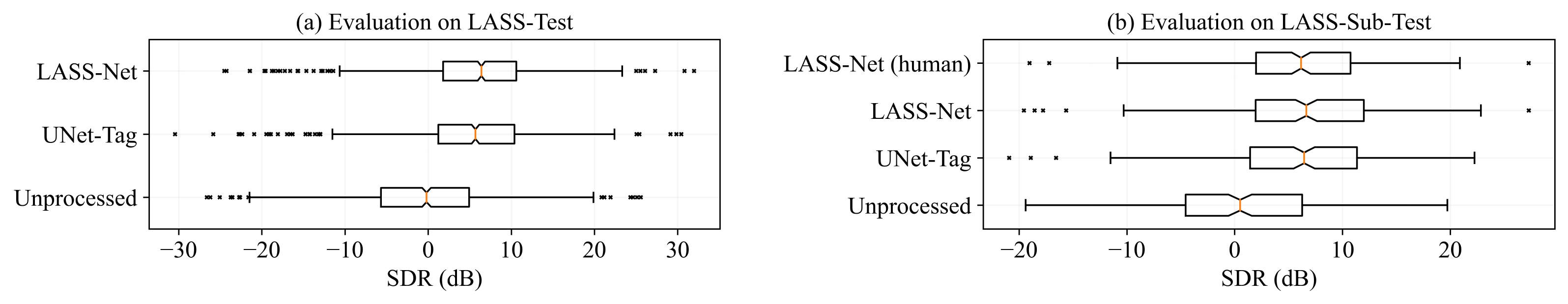}
  \caption{Box plots of the SDR scores on two experiments: (a) Evaluation on LASS-Test (b) Evaluation on LASS-Sub-Test. Red solid line represents the median value.}
  \label{fig-2}  
  \vspace{-1.5em}
\end{figure*}

\subsection{Data processing}
We load audio signals using \num{32} kHz sampling rate. STFT is calculated with a frame size of \num{1024} and a hop size of \num{512}. An audio clip of \num{10} seconds results in a spectrogram with shape of $513 \times 626$. We convert all language descriptions to lower case and remove punctuation. One special token ``\texttt{\textless SOS\textgreater}'' is added to the start of each sentence. We tokenize our language descriptions corpus using the WordPiece \cite{wu2016google} to match with the BERT pre-trained vocabulary ($\sim$\num{30}k).

\subsection{Baseline system}
Previous methods investigated separating sounds from a mixture given sound event tags \cite{kong2020source, ochiai2020listen}. We implement a sound event tags-queried source separation model as a baseline system. Specifically, the baseline model is queried by a multi-hot encoding of the sound event tags of the target source in AudioSet. To make a fair comparison between the separation performance queried by language descriptions and sound event tags, the architecture we used for the baseline model is similar to that of our proposed LASS-Net, except that the query network is a fully-connected layer with \num{256} nodes followed by a ReLU activation.

\subsection{Training procedure}
We set batch size to \num{16}. Adam \cite{kingma2014adam} optimizer is used for training with the learning rate of \num{3e-4}. We train the systems for \num{200000} iterations using four Nvidia-RTX-\num{3090}-\num{24}GB GPUs.

\subsection{Evaluation metrics}
We evaluate the model performance using four commonly used objective metrics \cite{vincent2006performance} in source separation: source to distortion ratio (SDR), source to inferences ratio (SIR), sources to artifact ratio (SAR) and scale-invariant SDR (SI-SDR) \cite{le2019sdr}. A higher number indicates a better performance.

\begin{table}[h]

\centering
\caption{Evaluation results on LASS-Test.}
\begin{tabular}[\linewidth]{c c c c c} 
 \hline
 Models & SDR & SIR & SAR & SI-SDR\\ 
 \hline
 Unprocessed & -0.46 & 12.17 & 0.55 & -0.53\\ 
 UNet-Tag & 5.40 & 16.28 & 4.75 & 4.24\\ 
 LASS-Net & 5.89 & 16.70 & 5.18 & 4.86\\ 
 \hline
\end{tabular}
\label{table-1}
% \vspace{-1.25em}
\end{table}

\begin{table}[h]
\centering
\caption{Evaluation results on LASS-Sub-Test.}
\begin{tabular}[\linewidth]{c c c c c} 
 \hline
Models & SDR & SIR & SAR & SI-SDR\\ 
 \hline
 Unprocessed & 0.56 & 12.27 & 1.68 & 0.53\\ 
 UNet-Tag & 6.18 & 16.41 & 5.29 & 4.90\\ 
 LASS-Net & 6.58 & 17.57 & 5.60 & 5.55\\ 
 LASS-Net (human) & 6.13 & 17.03 & 5.45 & 5.04\\ 
 \hline
\end{tabular}
\label{table-2}
\vspace{-1.25em}
\end{table}

\subsection{Results}
\subsubsection{Evaluation on LASS-Test}
\label{sec-4.3.1}
Three systems are evaluated on the LASS-Test in this experiment. The first method is the \textit{Unprocessed} system which directly uses the audio mixtures in the test set for evaluation. The second one is the UNet-based baseline system we implemented using sound event tags as queries, referred to as the \textit{UNet-Tags} system. The third one is the approach we proposed using AudioCaps language descriptions as queries, denotes as \textit{LASS-Net}. 

Table \ref{table-1} shows the experimental results and Figure \ref{fig-2} (a) depicts the box plot of the SDR scores. \textit{LASS-Net} achieves the SDR, SIR, SAR, SI-SDR improvement of \num{6.35} dB, \num{4.53} dB, \num{4.63} dB, and \num{5.39} dB, respectively, compared with the \textit{unprocessed} system. This proves the feasibility of LASS task and the effectiveness of our proposed approach. In addition, compared with the \textit{UNet-Tag} baseline system, \textit{LASS-Net} achieves the SDR, SIR, SAR, SI-SDR improvement of \num{0.49} dB, \num{0.42} dB, \num{0.43} dB, and \num{0.62} dB, respectively. Those performance improvements indicate that using natural language descriptions as queries leads to more accurate separation results, as compared with the use of sound event tags.

\subsubsection{Evaluation on LASS-Sub-Test}
Four systems are evaluated on the LASS-Sub-Test. The first three systems are \textit{Unprocessed}, \textit{UNet-Tag}, and \textit{LASS-Net}, respectively, which are same as those described in Section \ref{sec-4.3.1}. The fourth system is the model we proposed, but using our collected human description as queries, which we denote as \textit{LASS-Net (human)}. For \textit{LASS-Net (human)}, we average the evaluation results of five language queries for each target audio source.

Table \ref{table-2} shows the experimental results and Figure \ref{fig-2} (b) depicts the box plot of the SDR scores. \textit{LASS-Net (human)} achieves the SDR, SIR, SAR, SI-SDR improvement of \num{5.57} dB, \num{4.76} dB, \num{3.77} dB, and \num{4.51} dB, respectively, compared with the \textit{unprocessed} system. Although the human-annotated descriptions we collected have a different word distribution from that in AudioCaps, the performance of \textit{LASS-Net (human)} is only slightly lower as compared with the \textit{LASS-Net} system. In addition, \textit{LASS-Net (human)} outperforms \textit{UNet-Tag} baseline system on SIR, SAR and SI-SDR metrics by \num{0.62} dB, \num{0.16} dB, and \num{0.14} dB, respectively. These experimental results indicate the promising generalization performance of our proposed approach.

\begin{figure}[h]
  \centering
  \includegraphics[width=8cm]{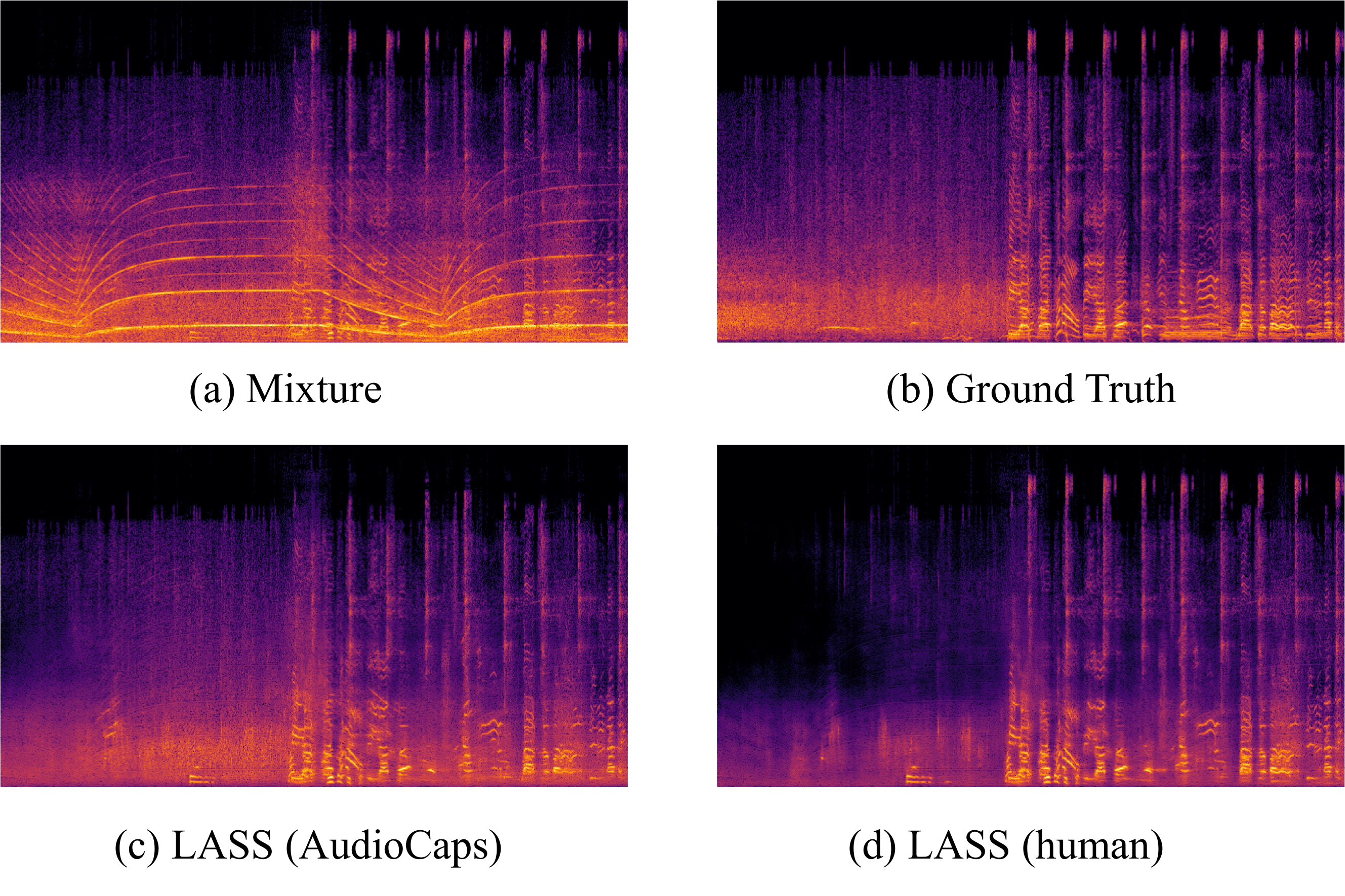}
  \caption{Spectrogram visualizations of (a) an audio mixture, (b) The ground truth target source, (c) The separated source queried by AudioCaps description: \textbf{``applauding followed by people singing and a tambourine''}, and (d) The separated source queried by our collected human description: \textbf{``a show start with audience applauding and then singing''}.}
  \label{fig-3}
  \vspace{-1.5em}
\end{figure}

We visualize the spectrograms of an audio mixture, the ground truth target source, the separated sources queried by AudioCaps description and our collected human description, respectively, in Figure \ref{fig-3}. Although these two language queries are different, we observe that the results of these two separations are similar and both are close to the ground truth source,  which is consistent with our experimental results. More audio separation samples are available on our GitHub page\footnote{\url{{https://github.com/liuxubo717/LASS}}}.
\section{Conclusions}
\label{sec-6}
We have presented a study of the LASS task, which, to our knowledge, is the first attempt bridging audio source separation with natural language queries. We presented a LASS-Net which separates target sources from mixtures using natural language queries. Experimental results show the promising separation results and generalization capabilities of LASS-Net. In future work, we will extend LASS to the scenario of complex acoustic mixtures and fine-grained natural language queries.
% with the state-of-the-art source separation methods \cite{tzinis2020improving, wisdom2021s}. 
\section{Acknowledgements}
This research was supported by a Newton Institutional Links Award from the British Council (Grant number 623805725), a Research Scholarship from the China Scholarship Council (CSC), and a PhD Studentship from the University of Surrey.

\bibliographystyle{IEEEtran}

\bibliography{mybib}

\end{document}